\documentclass{sig-alternate-2013}

\begin{document}
%
\conferenceinfo{PIR'15,Privacy-Preserving IR: When Information Retrieval Meets Privacy and Security, SIGIR 2015 Workshop} {August 13th, 2015, Santiago, Chile} 
\CopyrightYear{2015} 
\crdata{978-1-4503-2034-4/13/07} 
\clubpenalty=10000 
\widowpenalty = 10000

\title{Privacy-Preserving Multi-Document Summarization}
%
%
%
%
%

\numberofauthors{1} 
%
\author{
%
%
\alignauthor{Lu\'{i}s Marujo$^{1,2,3}$, Jos\'{e} Port\^{e}lo$^{2,3}$, Wang Ling$^{1,2,3}$, David Martins de Matos$^{2,3}$, Jo\~{a}o P. Neto$^{2,3}$, 
      Anatole Gershman$^{1}$, Jaime Carbonell$^{1}$, Isabel Trancoso$^{2,3}$, Bhiksha Raj$^{1}$ \\}
\affaddr{$^1$ Language Technologies Institute, Carnegie Mellon University, Pittsburgh, PA, USA \\
         $^2$ Instituto Superior T\'{e}cnico, Universidade de Lisboa, Lisbon, Portugal \\
         $^3$ INESC-ID, Lisbon, Portugal \\}
\email{ \{luis.marujo,jose.portelo,wlin,david.matos,joao.neto,isabel.trancoso\}@inesc-id.pt, \{anatoleg,jgc,bhiksha\}@cs.cmu.edu}
}

\maketitle
\begin{abstract}
State-of-the-art extractive multi-document summarization systems are usually designed without any concern about privacy issues, meaning that all documents are open to third parties. In this paper we propose a privacy-preserving approach to multi-document summarization. Our approach enables other parties to obtain summaries without learning anything else about the original documents' content.

We use a hashing scheme known as Secure Binary Embeddings to convert documents representation containing key phrases and bag-of-words into bit strings, allowing the computation of approximate distances, instead of exact ones. Our experiments indicate that our system yields similar results to its non-private counterpart on standard multi-docu-ment evaluation datasets.
\end{abstract}




\section{Introduction}
\label{sec:intro}


{\em Extractive Multi-document Summarization} (EMS) is the problem of extracting the most important sentences in a set of documents. State-of-the-art solutions for EMS based on {\em Waterfall KP-Centrality} achieve excellent results \cite{Marujo:SEM:2015}. A limitation to the usage of such methods is their assumption that the input texts are of public domain. However, problems arise when these documents cannot be made public. Consider the scenario where a company has millions of classified documents organized into several topics. The company may need to obtain a summary from each topic, but it lacks the computational power or know-how to do so. At the same time, they can not share those documents with a third party with such capabilities, as they may contain sensitive information. As a result, the company must {\em obfuscate} their own data before sending it to the third party, a requirement that is seemingly at odds with the objective of extracting summaries from it.

In this paper, we propose a new {\em privacy-preserving} technique for EMS based on Secure Binary Embeddings (SBE) \cite{SBE} that enables exactly this -- it provides a mechanism for obfuscating, not only named-entities \cite{Zhang:2014:PIR}, but the complete data, while still achieving near state-of-art performance in EMS. SBE is a kind of locality-sensitive hashing algorithm which converts data arrays such as bag-of-words vectors to obfuscated bit strings through a combination of random projections followed by banded quantization. The method has information theoretic guarantees of security, ensuring that the original data cannot be recovered from the bit strings.

They also provide a mechanism for computing distances between vectors that are close to one another without revealing the global geometry of the data, such as the number of features, consequently enabling tasks such as EMS. This is achievable because, unlike other hashing methods which require exact matches for performing retrieval or classification tasks, SBE allows for a near-exact matching: the hashes can be used to estimate the distances between vectors that are very close, but provably provide no information whatsoever about the distance between vectors that are farther apart.
The usefulness of SBE has already been shown in privacy-preserving important passage retrieval \cite{Marujo:PIR:2014} and speaker verification \cite{PPSV_SBE} systems, yielding promising results.


\section{Related Work}
\label{sec:related_work}

\subsection{Multi-document Summarization}

Most of the current work in automatic summarization focuses on extractive summarization. Popular baselines for multi-document summarization fall into one of the following general models: Centrality-based \cite{radev:et:al:2004,erkan:radev:2004,Ribeiro&Marujo:2013}, Maximal Marginal Relevance (MMR) \cite{carbonell:goldstein:1998,Guo:2010,Lim:2012}, 
and Coverage-based methods \cite{lin:hovy:2000}. 
Additionally, methods such as KP-Centrality \cite{Ribeiro&Marujo:2013}, which is centrality and coverage-based, follow more than one paradigm. In general, Centrality-based models are used to produce generic summaries, while the MMR family generates query-oriented ones. Coverage-based models produce summaries driven by words, topics or events.

We use the Waterfall KP-Centrality method because it is a state-of-the-art EMS method, but the ideas in this work could be applied to any other EMS methods.

\subsection{Privacy-Preserving Methods}
In this work, we focus on creating a method for performing EMS while keeping the original documents private. To the best of our knowledge, the combination of research lines has only been explored for the single-document summarization case \cite{Marujo:PIR:2014}. However, there are some additional recent works combining information retrieval and privacy. Most of these works use data encryption \cite{Jiang:2007,MurugesanEtAl:2010} 
to transfer the data in a secure way. The problem with these methods is that the entity responsible for producing the summaries will have access to the documents content, while our method guarantees that no party aside from the owner of the documents will have access to their content. Another secure information retrieval methodology is to obfuscate queries, which hides user topical intention but does not secure the content of the documents \cite{Pang:2012}.

In many areas, the interest in privacy-preserving methods where two or more parties are involved and they wish to jointly perform a given operation without disclosing their private information is not new, and several techniques such as Garbled Circuits (GC), 
Homomorphic Encryption (HE) 
and Locality-Sensitive Hashing (LSH) 
have been introduced. However, they all have limitations regarding the EMS task we wish to address. Until recently, GC methods were extremely inefficient and difficult to adapt, specially when the computation of non-linear operations, such as the cosine distance, is required. Systems based on HE techniques usually require extremely long amounts of time to evaluate any function of interest. The LSH technique allows for near-exact match detection between data points, but does not provide any actual notion of distance, leading to degradation of performance in some applications. As a result, we decided to consider SBE as the data privacy for our approach, as it does not show any of the disadvantages mentioned above for the task at hand. 

\section{Multi-document Summarization}
\label{sec:MDS}
To determine the most representative sentences of a set of documents, we used a multi-document approach based on KP-Centrality \cite{Ribeiro&Marujo:2013}. This method is adaptable and robust in the presence of noisy input. This is an important aspect since using several documents as input frequently increases the amount of unimportant content. 

Waterfall KP-Centrality iteratively combines the summa-ries of each document that was generated using KP-Centrali-ty following a cascade process: it starts by merging the intermediate summaries of the first two documents, according to their chronological order. This merged intermediate summary is then summarized and merged with the summary of following document. We iterate this process through all documents until the most recent one. The summarization method uses as input a set of key phrases that we extract from each input document, joins the extracted sets, and ranks the key phrases using their frequency. To generate each intermediate summary, we use the top key phrases, excluding the ones that do not occur in the input document.

KP-Centrality extracts a set of key phrases using a supervised approach 
and combines them with a bag-of-words model in a compact matrix representation, given by:
\begin{equation}
\begin{bmatrix}
  w(t_1,p_1) & \dots & w(t_1,p_N) & w(t_1,k_1) & \dots & w(t_1,k_M)\\
  \vdots & & & & & \vdots\\
  w(t_T,p_1) & \dots & w(t_T,p_N) & w(t_T,k_1) & \dots & w(t_T,k_M)\\
\end{bmatrix},
\label{eq:kpcentrality}
\end{equation}
where $w$ is a function of the number of occurrences of each term $t$ in every passage $p$ or key phrase $k$, $T$ is the number of terms, $N$ is the number of sentences and $M$ is the number of key phrases. Then, using $I \cup K \triangleq p_1,\dots,p_N,k_1,\dots,k_M=q_1,\dots,q_{N+M}$, a support set $S_{i}$ is computed for each passage $p_i$ using:
\begin{equation}
  S_i \triangleq \{s \in I \cup K : sim(s, q_i) > \varepsilon_i \wedge s \neq q_i\},
\label{eq:kpsupportset}
\end{equation}
for $i=1,\dots,N+M$. Passages are ranked excluding the set of key phrases ({\em artificial passages}) according to:
\begin{equation}
\operatorname*{arg\,max}_{s \in (\cup^{N}_{i=1}S_i)-K} \big|\{S_i: s \in S_i\}\big|.
\label{eq:kpmodel:centrality}
\end{equation}

A support set is a group of the most semantically related passages. These semantic passages are selected using heuristics based on the passage order method \cite{Ribeiro&Marujo:2013}. The metric that is normally used is the cosine distance.

\section{Secure Binary Embeddings}
\label{sec:SBE}
\begin{figure}[ht]
\centering
\includegraphics[scale=0.41]{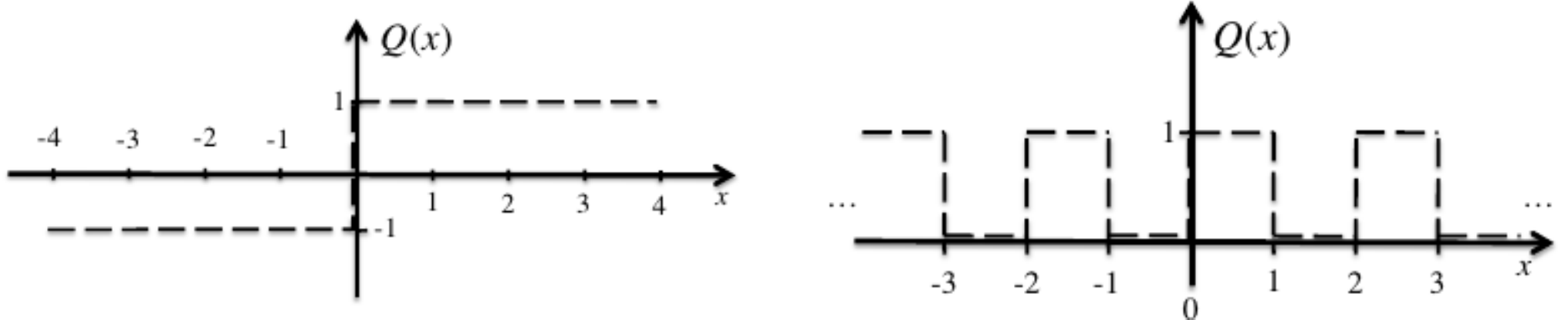}
\caption{1-bit quantization functions.}
\label{figure__1bit_quantization}
\end{figure}

An SBE is a scheme for converting vectors to bit sequences using quantized random projections. It produces a LSH method with an interesting property: if the Euclidean distance between two vectors is lower than a certain threshold, then the Hamming distance between their hashes is proportional to the Euclidean distance; otherwise, no information can be infered. This scheme is based on the concept of Universal Quantization (UQ)\cite{UQ}, which redefines scalar quantization by forcing the quantization function to have non-contiguous quantization regions. That is, the quantization process converts an $L$-dimensional vector ${\bf{x}} \in \mathbb{R}^{\it{L}}$ into an $M$-bit binary sequence, where the $m$-th bit is defined by:
\begin{equation}
q_{m}(\bf x) = Q \left ( \frac{\left < \bf{x},\bf{a}_{m} \right > + w_{m}}{\Delta_{m}} \right ) \label{eq:scalarsbe}
\end{equation}
Here $\left<,\right>$ represents a dot product.  $\bf{a}_m \in \mathbb{R}^{\it{L}}$ is a ``measurement'' vector comprising $L$ i.i.d. samples drawn from $\mathcal{N}(\mu=0,\sigma^2)$, $\Delta_m$ is a precision parameter and $w_m$ is random number drawn from a uniform distribution over $[0,\Delta_m]$. $Q(\cdot)$ is a quantization function given by $Q(x) = \lfloor x\%2 \rfloor$. We can represent the complete quantization into $M$ bits compactly in vector form:
\begin{equation}
{\bf{q}}(\bf x) = Q \left ( {\bf{\Delta}}^{-1} ( {\bf{Ax}}+{\bf{w}}) \right ) \label{eq:univquantizer}
\end{equation}
Here $\bf{q}(\bf x)$ is an $M$-bit binary vector, which we will refer to as the {\em hash} of $\bf{x}$,  ${\bf{A}} \in \mathbb{R}^{\it{M}\times \it{L}}$ is a matrix of random elements drawn from $\mathcal{N}(\mu=0,\sigma^{2})$, ${\bf{\Delta}}$ is a diagonal matrix with entries $\Delta_{m}$ and ${\bf{w}} \in \mathbb{R}^{\it{M}}$ is a vector of random elements drawn from a uniform distribution over $[0,\Delta_m]$. The universal 1-bit quantizer of Equation \ref{eq:scalarsbe} maps the real line onto $1/0$ in a banded manner, where each band is $\Delta_m$ wide.  Figure \ref{figure__1bit_quantization} compares conventional  scalar 1-bit quantization (left panel) with the equivalent universal 1-bit quantization (right panel).

The binary hash vector generated by the Universal Quantizer of Equation \ref{eq:univquantizer} has an interesting property: the hamming distance $Hamm(\bf{q}(\bf x),\bf{q})(\bf y))$ between the hashes of two vectors $\bf{x}$ and $\bf {y}$ is correlated to the Euclidean distance $\|\bf{x}-\bf{y}\|$ between the two vectors, if the Euclidean distance between the two vectors is less than a threshold (which depends on $\Delta_m$). However, if the distance between $\bf{x}$ and $\bf{y}$ is greater than this threshold, $Hamm(\bf{q}(\bf x),\bf{q})(\bf y))$ yields no information about the true distance between the vectors \cite{SBE}.

\begin{figure}[ht]
\centering
\includegraphics[scale=0.28]{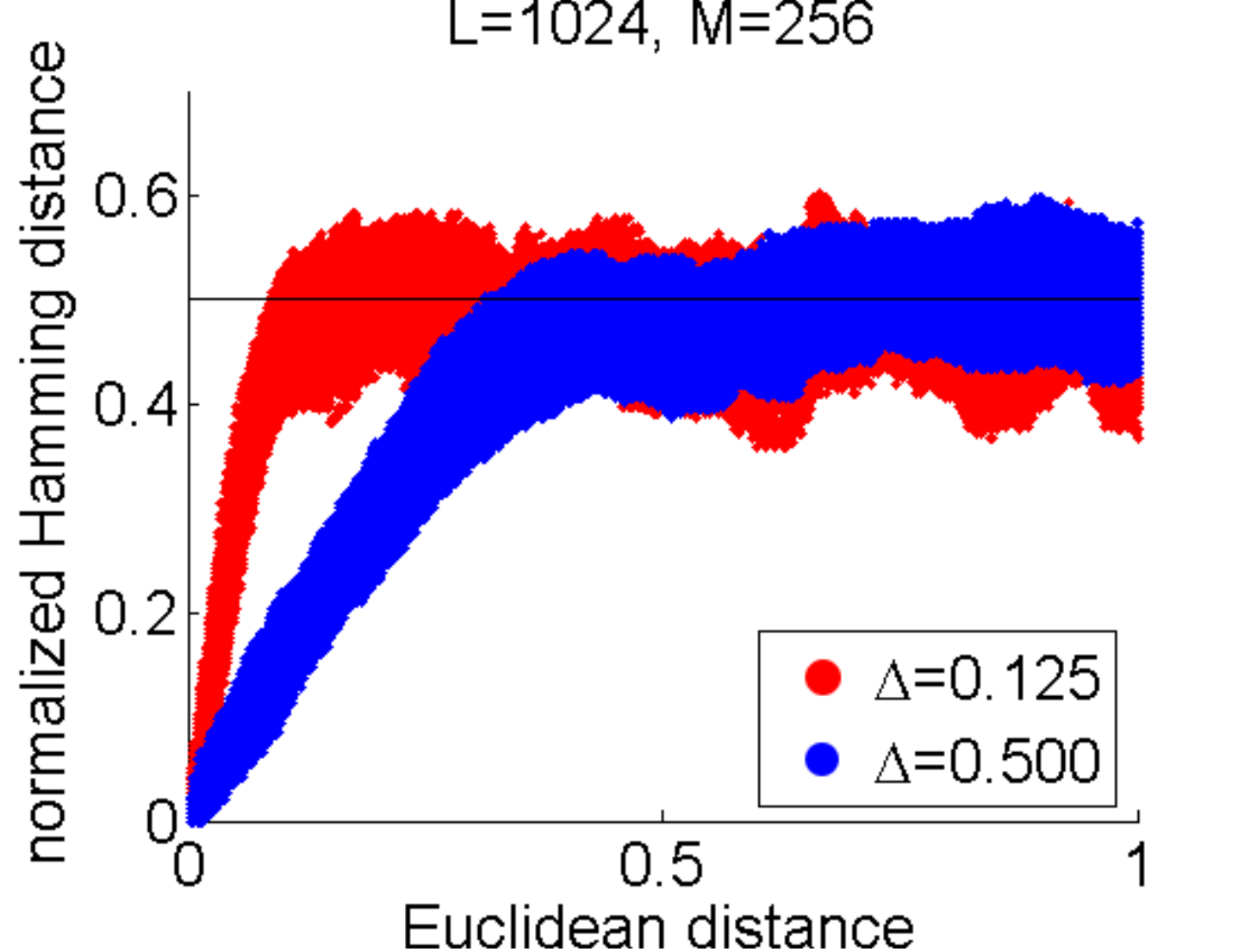}
\includegraphics[scale=0.28]{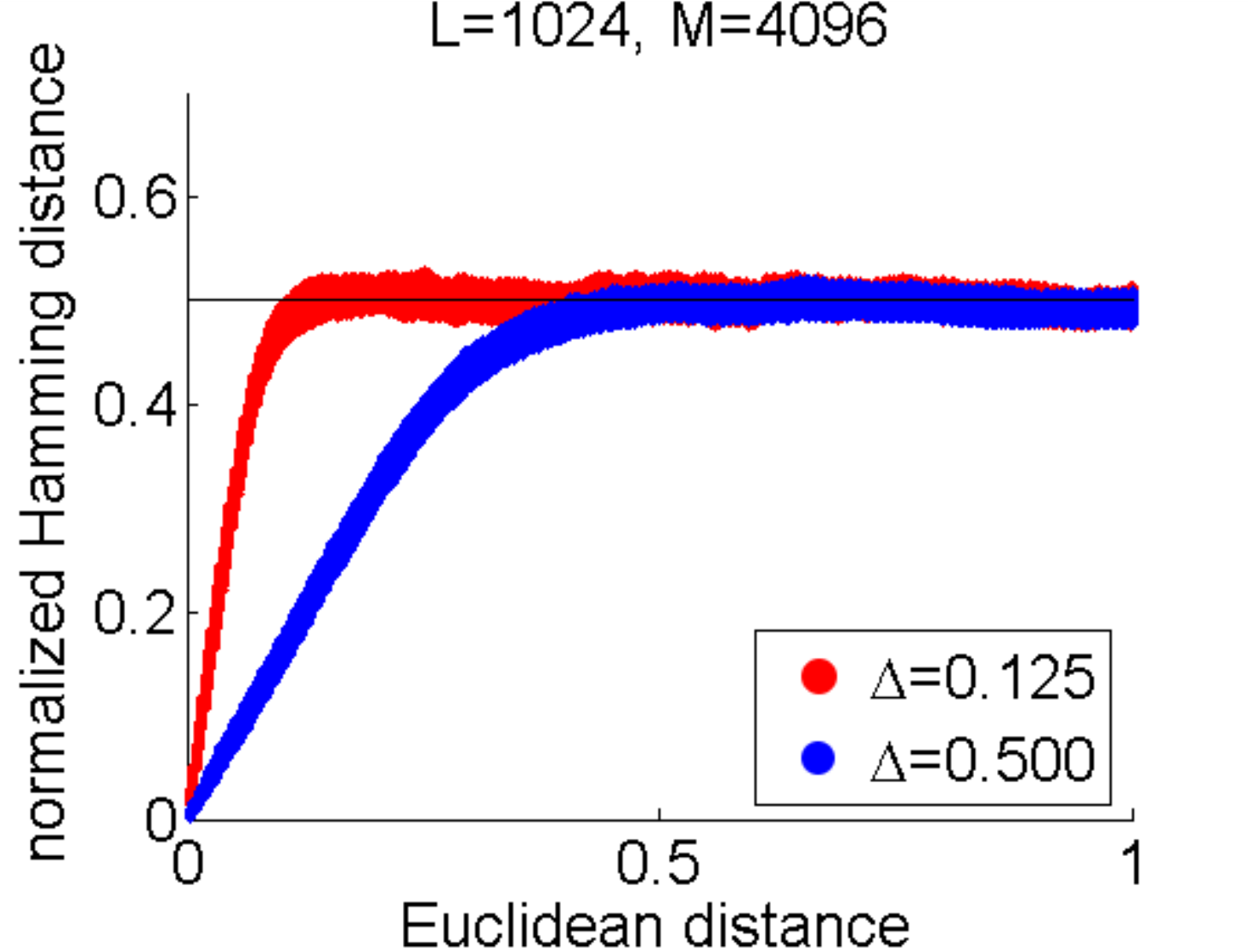}
\caption{Embedding behaviour for different values of $\Delta$ and different amounts of measurements $M$.}
\label{figure__sbe}
\end{figure}

In order to illustrate how this scheme works, we randomly generated samples in a high-dimensional space ($L=1024$) and plotted the normalized Hamming distance between their hashes against the Euclidean distance between the respective samples. This is presented in Figure \ref{figure__sbe}. The number of bits in the hash is also shown in the figures.

We note that in all cases, once the normalized distance exceeds $\Delta$, the hamming distance between the hashes of two vectors ceases to provide any information about the true distance between the vectors. We will find this properly useful in developing our privacy-preserving MDS system.

We also see that changing the value of the precision parameter $\Delta$ allows us to adjust the distance threshold until which the hamming distance is informative. Also, increasing the number of bits $M$ leads to a reduction of the variance of the Hamming distance.
Yet another interesting property conjectured for the SBE is that recovering $\bf{x}$ from $\bf{q}(\bf x)$ is $NP$-hard, even given $\bf A$.

\section{Secure Multi-document \\ Summarization}
\label{sec:secure_multidocument_summarization}

Our methodology consists in iteratively running the secure single-document summarization method \cite{Marujo:PIR:2014}, which comprises four stages. In the first stage we obtain a representation of each document, which is the first step of the KP-Centrality method. In the second stage we compute SBE hashes using the document representation. The third stage ranks the passages, which corresponds to the second step of the KP-Centrality method. Because we are now working with SBE hashes instead of the original document representation, this is performed using the Hamming distance instead of the cosine distance. Finally, the last stage is to use the ranks of sentences to obtain the summary.

Our approach for a privacy-preserving multi-document summarization system closely follows the formulation presented in Section \ref{sec:MDS}. However, there is a very important difference in terms of who performs each of the steps of the single-document summarization method. Typically, the only party involved, Alice, who owns the original documents, performs key phrase extraction, combines them with the bag-of-words model in a compact matrix representation, computes the support sets for each document and finally uses them to retrieve the summaries. In our scenario, Alice does not know how to extract the important passages from the document collection and/or does not possess the computational power to do so. Therefore, she must outsource the summarization process to a another entity, Bob, who has these capabilities. However, Alice must first obfuscate the information contained in the compact matrix representation. If Bob receives this information as is, he could use the term frequencies to infer on the contents of the original documents and gain access to private or classified information Alice does not wish to disclose. Alice computes binary hashes of her compact matrix representation using the method described in Section \ref{sec:SBE}, keeping the randomization parameters ${\bf A}$ and ${\bf w}$ to herself. She sends these hashes to Bob, who computes the support sets and extracts the important passages. Because Bob receives binary hashes instead of the original matrix representation, he must use the normalized Hamming distance instead of the cosine distance in this step, since it is the metric the SBE hashes best relate to. Finally, he returns the hashes corresponding to the important passages to Alice, who then uses them to get the information she desires. These steps are repeated as many times as needed until the multi-document summarization process is complete.

\section{Experiments and Results}
\label{sec:experiments}
In this section we illustrate the performance of our privacy-preserving approach to EMS and how it compares to its non-private counterpart. We start by presenting the datasets we used in our experiments, then we describe the experimental setup and finally we present some results.

To assess the quality of the summaries generated by our methods, we used ROUGE--1 \cite{lin:2004} on DUC 2007 and TAC 2009 datasets. DUC 2007\footnote{\small{http://www-nlpir.nist.gov/projects/duc/duc2007/tasks.html}} dataset includes 45 clusters of 25 newswire documents and 4 human-created 250-word reference summaries. TAC 2009\footnote{\small{http://www.nist.gov/tac/2009/Summarization/}} has 44 topic clusters. Each topic has 2 sets of 10 news documents. There are 4 human-created 100-word reference summaries for each set. The reference summaries for the first set are query-oriented and for the second set are update
summaries. In this work, we used the first set of reference summaries. We evaluated the different models by generating summaries with 250 words.

We present some baseline experiments in order to obtain reference values for our approach. We generated 250 words summaries for both TAC 2009 and DUC 2007 datasets. For both experiments, we used the cosine and the Euclidean distance as evaluation metrics, since the first is the usual metric for computing textual similarity, but the second is the one that relates to the Secured Binary Embeddings technique. All results are presented in terms of ROUGE \cite{lin:2004}, in particular ROUGE--1, which is the most widely used evaluation measure for this scenario. The results we obtained for the TAC 2009 and the DUC 2007 are presented in Table \ref{table:Default_KP_CentralityResults}.
\begin{table}[!t]
\center
\begin{tabular}{l|c c}
Metric             & TAC 2009 & DUC 2007 \\\hline
Cosine distance    & 0.514 & 0.370   \\
Euclidean distance & 0.489 & 0.364   \\
\end{tabular}
\caption{Reference Waterfall KP-Centrality results with 40 key phrases, in terms of ROUGE--1.}
\label{table:Default_KP_CentralityResults}
\end{table}

We considered 40 key phrases in our experiments since it is the usual choice when news articles are considered \cite{Ribeiro&Marujo:2013}. 
As expected, we notice some slight degradation when the Euclidean distance is considered, but we still achieve better results than other state-of-the-art methods such as MEAD \cite{radev:et:al:2004}, MMR \cite{carbonell:goldstein:1998}, Expect n-call@k \cite{Lim:2012}, and LexRank \cite{erkan:radev:2004}.

Reported results in the literature include 
$\text{ROUGE--1}=0.328$ and $0.415$ using MEAD,
$\text{ROUGE--1}=0.327$ and $0.392$ using MMR, 
$\text{ROUGE--1}=0.321$ and $0.387$ using Expect n-call@k for the DUC 2007 and TAC 2009 datasets, respectively \cite{Marujo:SEM:2015}. This means that the forced change of metric due to the intrinsic properties of SBE and the multiple application of SBE does not affect the validity of our approach in any way.

For our privacy-preserving approach we performed experiments using different values for the SBE parameters. The results we obtained in terms of ROUGE for the DUC 2007 and the TAC 2009 datasets are presented in Tables \ref{table:KP-CentralityResults_with_privacy_DUC2007} and \ref{table:KP-CentralityResults_with_privacy_TAC2009}, respectively.
\begin{table}[!t]
\center
\begin{tabular}{l|ccccc}
leakage      & $\sim5\%$ & $\sim25\%$ & $\sim50\%$ & $\sim75\%$ & $\sim95\%$ \\
\hline
{\em bpc}=4  & 0.331     & 0.343     & 0.338      & 0.347      & 0.347      \\
{\em bpc}=8  & 0.339     & 0.341     & 0.341      & 0.352      & 0.356      \\
{\em bpc}=16 & 0.336     & 0.348     & 0.337      & 0.350      & 0.351      \\
\end{tabular}
\caption{Waterfall KP-Centrality using SBE and the DUC 2007 corpus, in terms of ROUGE--1.}
\label{table:KP-CentralityResults_with_privacy_DUC2007}
\end{table}
\begin{table}[!t]
\center
\begin{tabular}{l|ccccc}
leakage      & $\sim5\%$ & $\sim25\%$ & $\sim50\%$ & $\sim75\%$ & $\sim95\%$ \\
\hline
{\em bpc}=4  & 0.475     & 0.472     & 0.458      & 0.478      & 0.487      \\
{\em bpc}=8  & 0.462     & 0.472     & 0.469      & 0.473      & 0.486      \\
{\em bpc}=16 & 0.448     & 0.467     & 0.462      & 0.484      & 0.491      \\
\end{tabular}
\caption{Waterfall KP-Centrality using SBE and the TAC 2009 corpus, in terms of ROUGE--1.}
\label{table:KP-CentralityResults_with_privacy_TAC2009}
\end{table}
Leakage denotes the percentage of SBE hashes that the normalized Hamming distance $d_{H}$ is proportional to the Euclidean distance $d_{E}$ between the original data vectors. The amount of leakage is controlled by $\Delta$. Bits per coefficient ($bpc$) is the ratio between the number of measurements $M$ and the dimensionality of the original data vectors $L$, i.e., $bpc=M/L$. Unsurprisingly, increasing the amount of leakage (i.e., increasing $\Delta$) leads to improving the summarization results. However, changing $bpc$ does not lead to improved performance. The reason for this might be due to the Waterfall KP-Centrality method using support sets that consider multiple partial representations of all documents. Even so, the most significant results is that for $95\%$ leakage there is an almost negligible loss of performance. This scenario, however, does not violate our privacy requisites in any way, since although most of the distances between hashes are known, it is not possible to use this information to obtain anything about the original information.

%
%
%
\section{Conclusions and Future Work}
\label{sec:conclusions}
In this work, we introduced a privacy-preserving technique for performing Extractive Multi-document Summarization that has similar performance to their non-private counterpart. Our Secure Binary Embeddings based approach provides secure multiple documents representations that allows for sensitive information to be processed by third parties without any risk of sensitive information disclosure. We also found it rather interesting to observe such a small degradation on the results given that we needed to compute SBE hashes on each iteration of our algorithm.

Future work will explore the possibility of having multiple rather than a single entity supplying all the documents.

\section{Acknowledgments}
We would like to thank FCT for supporting this research through grants UID/CEC/50021/2013, PTDC/EIA-CCO/ 122542/2010, CMUP-EPB/TIC/0026/2013, and CMU-Portugal. 

%

\bibliographystyle{abbrv}
\bibliography{sigirrsp098-v2}  
%
%
%
\end{document}